\documentclass[conference]{IEEEtran}
\IEEEoverridecommandlockouts
\usepackage{cite}
\usepackage{xcolor}
\usepackage{indentfirst}
\newif\ifwithappendix
\withappendixtrue      % 完整版用 true
\newcommand{\appnote}[1]{\ifwithappendix\footnote{#1}\fi}

\usepackage{newtxmath}
\usepackage{graphicx}
\usepackage{booktabs}
\usepackage{multirow}
\usepackage{enumitem}
\usepackage{tabularx}
\usepackage{ragged2e}
\usepackage{array}
\usepackage{textcomp}
\usepackage{algorithm}
\usepackage{algorithmic}
\usepackage{makecell}
\usepackage{amsmath,amsfonts}
\usepackage{url}
\usepackage{xurl}
\usepackage[hidelinks]{hyperref}
\usepackage{balance}

\def\BibTeX{{\rm B\kern-.05em{\sc i\kern-.025em b}\kern-.08em
    T\kern-.1667em\lower.7ex\hbox{E}\kern-.125emX}}

\providecommand{\Description}[1]{}

\begin{document}

\title{Proactive Detection of GUI Defects in Multi-Window Scenarios via Multimodal Reasoning}

\author{
\begin{tabular}{c}

\begin{tabular}{cc}
\begin{tabular}[t]{c}
\textbf{Xinyao Zhang, Rui Wang, Jinhao Cui, Haotian Huang}\\
\textit{School of Computer Science and Artificial Intelligence}\\
\textit{Wuhan University of Technology}\\
Wuhan, China\\
{xinyaozhang, 339993, cuiwut0123, huanghaotian}@whut.edu.cn
\end{tabular}
&
\hspace{2em}
\begin{tabular}[t]{c}
\textbf{Wei Xue}\\
\textit{Dongfeng Motor Corporation}\\
Wuhan, China\\
xuewei@dfmc.com.cn
\end{tabular}
\end{tabular}

\\[1.2em]

\begin{tabular}[t]{c}
\textbf{Wenhua Hu, Jianwen Xiang, Rui Hao\textsuperscript{*}}\\
\textit{School of Computer Science and Artificial Intelligence}\\
\textit{Engineering Research Center of Transportation Information and Safety (ERCTIS), MoE of China}\\
\textit{Wuhan University of Technology}\\
Wuhan, China\\
{whu10, jwxiang, ruihao}@whut.edu.cn\\[-1.0em]
\end{tabular}

\end{tabular}
}
\maketitle
\vspace{-24pt}

\begin{abstract}
% Multi-window scenarios, such as split-screen and foldable modes, require mobile applications to adapt to changing display space and dynamic layout reflow, making display defects such as text truncation, widget occlusion, and layout overlap more likely to occur. Existing GUI defect detection methods mainly rely on manual inspection, heuristic rules, or purely visual classification, limiting their ability to support both dynamic scenario triggering and semantic-level defect identification. To address this limitation, we propose an end-to-end framework for software interface defect detection that combines visual marking with chain-of-thought prompting. The framework extends DroidBot to trigger split-screen, foldable, and window-transition scenarios and collect multimodal evidence, uses Set-of-Mark (SoM) for widget-level visual-semantic alignment, and leverages multimodal large language models for defect detection, localization, and explanation. Experiments on 50 real-world Android applications show that split/fold settings substantially increase the exposure of layout-related defects, with text truncation increasing by 184\% compared with the conventional setting. At the application level, our method identifies 40 defect apps under split/fold settings, with an FPR of 10.00\% and an FNR of 11.11\%, outperforming OwlEye and YOLO. At the fine-grained level, it achieves the best F1 score of 87.2\% for widget occlusion detection.

Multi-window mobile scenarios, such as split-screen and foldable modes, make GUI display defects more likely by forcing applications to adapt to changing window sizes and dynamic layout reflow. Existing detection techniques are limited in two ways: they are largely passive, analyzing screenshots only after problematic states have been reached, and they are mainly designed for conventional full-screen interfaces, making them less effective in multi-window settings.
We propose an end-to-end framework for GUI display defect detection in multi-window mobile scenarios. The framework proactively triggers split-screen, foldable, and window-transition states during app exploration, uses Set-of-Mark (SoM) to align screenshots with widget-level interface elements, and leverages multimodal large language models with chain-of-thought prompting to detect, localize, and explain display defects. We also construct a benchmark of GUI display defects using 50 real-world Android applications.
Experimental results show that multi-window settings substantially increase the exposure of layout-related defects, with text truncation increasing by 184\% compared with conventional full-screen settings. At the application level, our method detects 40 defect-prone apps with a false positive rate of 10.00\% and a false negative rate of 11.11\%, outperforming OwlEye and YOLO-based baselines. At the fine-grained level, it achieves the best F1 score of 87.2\% for widget occlusion detection.
\end{abstract}


\section{Introduction}
% With the growing adoption of foldable devices and split-screen multitasking, mobile applications no longer operate only in the conventional full-screen smartphone setting. Instead, they must adapt to changing window modes and reduced display space \cite{android_foldables,android_multiwindow,android_screen_compatibility}. Under such dynamic and space-constrained layouts, display defects that are less visible in full-screen mode are often amplified, including text truncation, widget occlusion, layout overlap, and conflicts with the system UI, which directly harm readability, operability, and user experience \cite{nie2024sok,su2022metamorphosis,liu2020owl}.

Mobile applications are increasingly used beyond the conventional full-screen smartphone setting. With the rapid adoption of foldable devices, tablets, split-screen multitasking, and freeform windows, modern apps must continuously adapt their interfaces to changing window sizes, aspect ratios, and display modes \cite{android_foldables,android_multiwindow,android_screen_compatibility}. Such adaptations are error-prone: when the available display area shrinks or changes dynamically, defects such as text truncation, widget occlusion, layout overlap, and conflicts with system UI elements become more likely to occur and more visible to users. These defects directly degrade readability, operability, and overall user experience \cite{nie2024sok,su2022metamorphosis,liu2020owl}.

% Graphical user interface (GUI) display defect detection in these scenarios faces two challenges. First, \emph{triggering such defects is difficult}. Many adaptation defects emerge only under specific window-mode changes, while traditional full-screen exploration cannot systematically cover split-screen, foldable, and resizing scenarios \cite{android_multiwindow,li2017droidbot}. Second, \emph{identifying such defects is difficult}. Phenomena that visually resemble ``occlusion'' or ``overlap'' may in fact be acceptable design choices, such as badge counters or floating buttons partially covering edge content. Methods based only on heuristic rules, static layout cues, or purely visual classifiers often lack sufficient understanding of interface semantics and design intent \cite{liu2020owl,su2022metamorphosis,nie2024sok}, making it difficult to distinguish \emph{acceptable occlusion} from \emph{erroneous occlusion}.

Existing GUI display defect detection techniques, however, are not designed for this new usage context. Prior approaches are largely \emph{passive}: they take already-captured screenshots as input and determine whether a display defect exists \cite{liu2020owl,su2022metamorphosis}. While effective for post hoc inspection, such methods cannot proactively steer app exploration toward defect-prone interface states. Moreover, existing studies, benchmarks, and detection models mainly focus on conventional \emph{full-screen} smartphone UIs, with limited support for split-screen, foldable, and other dynamic multi-window scenarios \cite{nie2024sok}. 
% As a result, current techniques are both \emph{passive} and \emph{full-screen-centric}, leaving a critical gap in detecting display defects under realistic multitasking usage.

Detecting display defects in multi-window environments is challenging for two reasons. First, \emph{triggering defects is difficult}. Many adaptation-related defects manifest only when an app undergoes specific window transitions, such as entering split-screen mode, resizing to a smaller region, or switching display posture on foldable devices. Traditional exploration tools, which primarily traverse apps under a fixed full-screen configuration, cannot systematically cover such scenarios \cite{li2017droidbot}. Second, \emph{identifying defects is difficult}. Multi-window interfaces often contain visually ambiguous patterns: apparent overlap or partial occlusion may either indicate a true defect or reflect intentional design choices, such as floating action buttons, badge counters, or transient overlays. Therefore, methods based only on heuristic rules, layout metadata, or purely visual classification often lack the semantic understanding needed to distinguish acceptable UI compositions from genuine display defects \cite{liu2020owl,su2022metamorphosis,nie2024sok}.

% To address these challenges, we propose an end-to-end framework for interface display defect detection in split-screen and foldable scenarios. The framework extends DroidBot \cite{li2017droidbot} to actively trigger split-screen, foldable, and window-transition scenarios while collecting aligned evidence. It further adopts Set-of-Mark (SoM) \cite{yang2023set} for widget-level visual-semantic alignment and combines multimodal large language models with chain-of-thought prompting (CoT) \cite{wei2022chain} for defect detection, localization, and explanation.

In this paper, we propose an end-to-end framework for GUI display defect detection in split-screen, foldable, and dynamic window-transition scenarios. Our key idea is to integrate \emph{proactive defect-triggering exploration} with \emph{multimodal defect reasoning}. On the exploration side, we extend DroidBot \cite{li2017droidbot} to actively induce split-screen, foldable, and window-resizing scenarios during app traversal, thereby increasing the likelihood of reaching defect-prone states. On the analysis side, we use Set-of-Mark (SoM) \cite{yang2023set} to align screenshots with widget-level interface elements, and then employ multimodal large language models with chain-of-thought prompting (CoT) \cite{wei2022chain} to reason about defect existence, localization, and explanation. In this way, the framework goes beyond passive screenshot inspection and supports active discovery of display defects in realistic multi-window usage settings.

\begin{figure}[t]
  \centering
  \includegraphics[width=\linewidth]{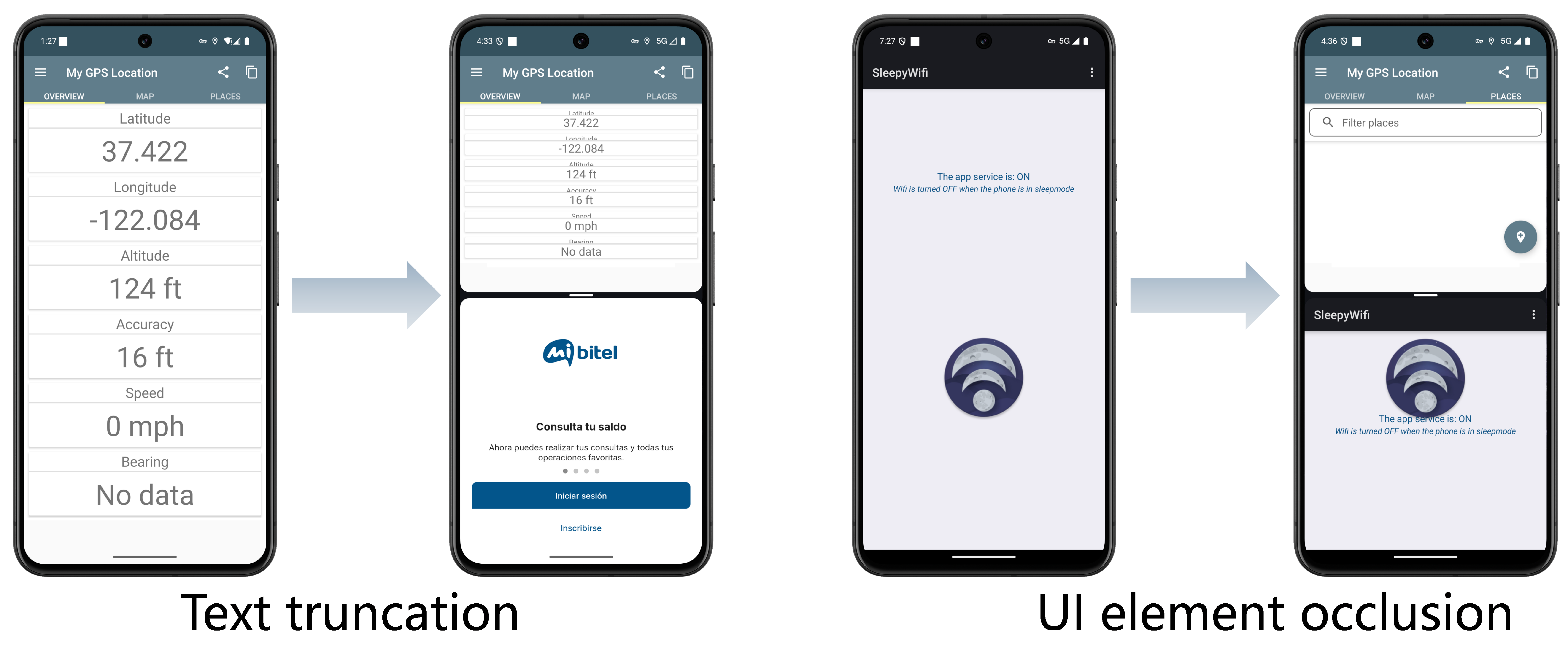}
  \caption{Examples of amplified display defects in split-screen and foldable scenarios.}
  \label{fig:intro-example}
\end{figure}

% The main contributions of this paper are as follows:
% \begin{itemize}[leftmargin=1.2em, itemsep=2pt, topsep=2pt]
%   \item \textbf{A multi-window triggering and evidence collection method.} By extending DroidBot, the method systematically triggers split-screen, foldable, and window-transition scenarios and collects aligned screenshot--widget tree--context data, improving the exposure of adaptation defects.
%   \item \textbf{A semantics-enhanced defect identification method based on SoM and CoT.} The method uses SoM for widget-level visual-semantic alignment and CoT-driven multimodal reasoning for explainable defect detection and widget-level localization.
%   \item \textbf{An empirical evaluation on real-world Android applications.} Results show that split-screen and foldable scenarios expose more layout-related defects and that the proposed method outperforms representative baselines in both overall detection and fine-grained defect recognition.
% \end{itemize}
The main contributions of this paper are as follows:
\begin{itemize}
    \item We identify a new testing gap in GUI display defect detection: existing methods are predominantly reactive, screenshot-based, and centered on full-screen smartphone UIs, while modern multi-window mobile environments require both active defect triggering and context-aware defect interpretation.
    \item We present an exploration framework that extends DroidBot to systematically trigger defect-prone split-screen, foldable, and window-transition scenarios during dynamic app traversal.
    \item We develop a multimodal defect detection approach that combines SoM-based widget grounding with MLLM reasoning to detect, localize, and explain display defects in complex multi-window interfaces.
    \item We construct, to the best of our knowledge, the first benchmark of GUI display defects in multi-window mobile scenarios using 50 Android apps, and use it to evaluate the proposed approach against existing baselines.
\end{itemize}

\section{Related Work}
\label{sec:related_work}

Research on mobile GUI quality assurance can be broadly grouped into three lines: (\emph{i}) heuristic- or rule-based usability evaluation and interface inspection; (\emph{ii}) deep learning-based GUI exploration and defect recognition; and (\emph{iii}) LLM/MLLM-based script generation and interaction decision-making. Different from conventional full-screen settings, this work focuses on display and adaptation defects caused by window mode changes, especially split-screen and foldable states, where both dynamic scenario triggering and semantic-level defect identification are required.

\subsection{Traditional GUI Testing Methods}
Traditional studies mainly rely on heuristic evaluation, remote usability testing, and interaction log analysis. Representative work includes Scholtz’s remote usability framework~\cite{scholtz2001adaptation}, mobile-oriented data collection platforms~\cite{liang2011remote}, mobile usability and gesture heuristics~\cite{machado2013heuristics,humayoun2017heuristics,bashir2019euhsa}, and GUI trace visualization tools for reducing manual analysis effort~\cite{jeong2020gui}. However, these methods depend heavily on predefined rules and thresholds, making them less robust to cross-application variation, resolution differences, and dynamic layout changes in split-screen and foldable scenarios. They also lack explicit modeling of interface semantics, and thus struggle to distinguish acceptable occlusion (e.g., badges or floating action buttons) from true defects.

\subsection{Deep Learning-based GUI Testing}
Deep learning has improved GUI exploration and defect recognition. Deep GUI predicts touch targets by heatmap regression to improve exploration on sparse interfaces~\cite{yazdanibanafshedaragh2021deep}. Humanoid and MUBot learn action strategies from human interaction traces~\cite{li2019humanoid,peng2022mubot}, while reinforcement learning has also been introduced to improve coverage and fault detection~\cite{pan2020reinforcement,romdhana2022deep}. For display defect detection, OwlEye applies convolutional neural network (CNN)-based modeling to identify multiple defect types from screenshots~\cite{liu2020owl}. Nevertheless, defect annotations are scarce and cross-app distribution shifts are substantial, limiting generalization. More importantly, purely visual methods do not explicitly model widget semantics or design intent, and therefore remain prone to semantically dependent misclassification.

\subsection{LLM-based GUI Testing}
Large language models (LLMs) further extend GUI testing to script generation and interaction decision-making. GPTDroid formulates testing as a question-answering-based decision process~\cite{liu2023chatting}; InputBlaster generates anomalous inputs for text widgets to improve crash discovery~\cite{liu2024testing}; and DroidBot-GPT textualizes GUI states and candidate actions for LLM-based action selection~\cite{wen2023droidbot}. However, existing studies mainly target functional testing rather than screenshot-level display defect detection. They usually rely on textualized GUI states, but lack explicit detection objectives and explainable oracles for visual defects. Although multimodal large language models (MLLMs) offer the potential to jointly use visual evidence and widget semantics, practical issues such as grounding stability, reproducibility, and reasoning cost remain.

\section{Approach}
\label{sec:approach}

\begin{figure*}[t]
  \centering
  \includegraphics[width=\textwidth]{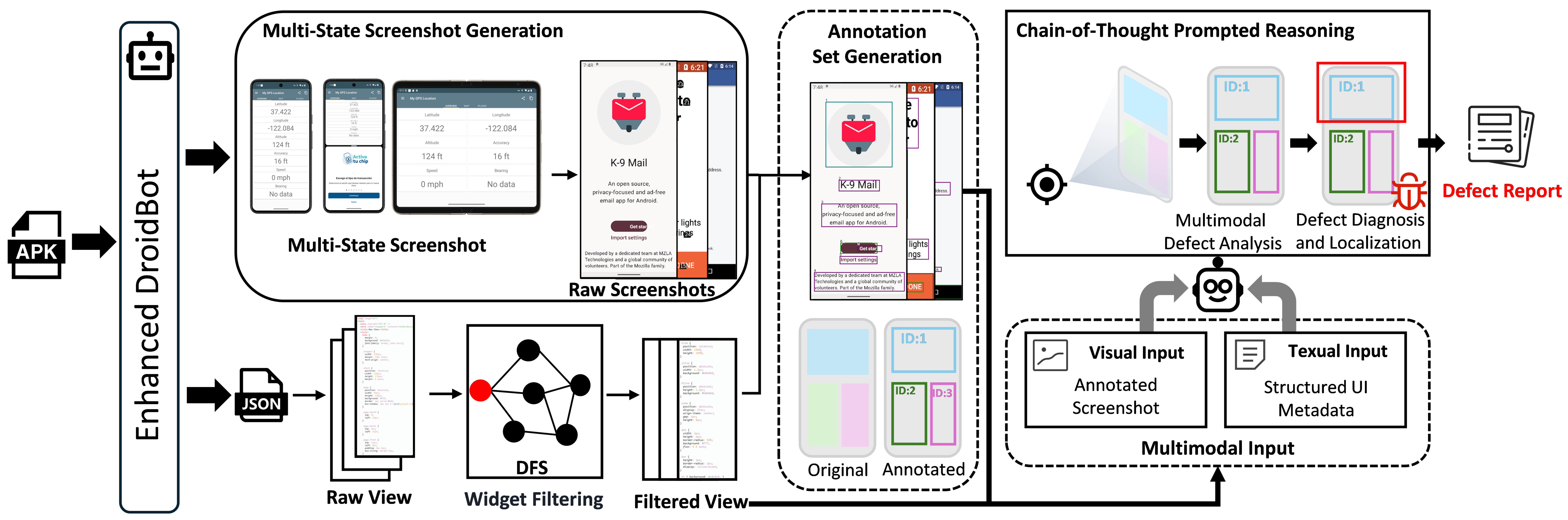}
  \caption{Framework overview of our approach. Enhanced DroidBot collects screenshots and view hierarchies across multiple window states, and the resulting multimodal inputs are analyzed with CoT-guided reasoning for defect diagnosis, localization, and structured reporting.}
  \label{fig:framework}
  \vspace{-4mm}
\end{figure*}

\subsection{Method Overview}
As shown in Fig.~\ref{fig:framework}, we propose an end-to-end framework for interface display defect detection in split-screen, foldable, and window-resizing scenarios. The framework addresses dynamic scenario triggering and semantic-aware defect identification in four stages. First, an enhanced DroidBot performs split-screen, foldable, and window-transition operations during exploration and collects screenshots, widget trees, and runtime context. Second, key widgets are filtered from the raw widget tree and serialized into structured user interface (UI) metadata. Third, a Set-of-Mark (SoM) representation is constructed to align screenshot regions with widget semantics. Finally, the marked screenshots and structured UI metadata are fed into a multimodal large language model with chain-of-thought prompting for defect diagnosis, localization, and explanation.

\subsection{Stage I: Multi-State Screenshot Generation and Evidence Collection}
DroidBot~\cite{li2017droidbot} is a model-based Android testing framework that extracts GUI information during exploration and builds a state model for event selection and path planning. It can also synchronously collect reproducible evidence, including GUI screenshots and the UI hierarchy. In our framework, we extend DroidBot with split-screen, foldable, and window-transition operations to support multi-window exploration and evidence collection.
To expose defects that are difficult to trigger in conventional full-screen scenarios, we extend DroidBot to perform four concrete window operations during exploration, namely Split Test, Fold Test, Drag Up, and Drag Down. These operations cover three representative transition categories: entering split-screen mode and adjusting the split ratio, switching foldable states, and resizing the window by dragging its boundaries. For each stable interface state, we collect an aligned evidence triplet:
\begin{equation}
\langle I,\; U,\; \Gamma \rangle,
\label{eq:triplet}
\end{equation}
where $I$ is the raw screenshot, $U$ is the UI hierarchy (widget-tree JSON), and $\Gamma$ is the runtime context, including the app, Activity, window-mode parameters, and timestamp. This triplet is the unified input for subsequent filtering, alignment, and defect reasoning.\appnote{Detailed implementation of the enhanced multi-window testing framework is provided in Appendix~\ref{app:droidbot}.}

\subsection{Stage II: Key Widget Filtering and Ordering}
The widget tree often contains many decorative or container nodes. Directly using it as input introduces noise and increases reasoning cost. We therefore perform lightweight widget abstraction and filtering, and then derive an approximate reading order.

\subsubsection{Widget Abstraction}
From $U$, we parse a widget set $\mathcal{C}=\{c_i\}_{i=1}^{n}$, where each widget is represented as
\begin{equation}
c_i = (id_i,\; type_i,\; text_i,\; b_i,\; click_i,\; rid_i),
\label{eq:control}
\end{equation}
where $b_i=(x_{1i},y_{1i},x_{2i},y_{2i})$ is the bounding box, $click_i$ is runtime clickability, and $rid_i$ is the resource identifier.

\subsubsection{Lightweight Filtering}
To remove nodes irrelevant to defect identification, we define a binary decision function $\varphi(\cdot)$ that preserves only interactive or semantically meaningful widgets while filtering out extremely small elements, decorative backgrounds, and empty containers:
\begin{equation}
\varphi(c_i)=\mathbb{I}\Big[
\operatorname{area}(b_i)\ge \tau_a \;\wedge\;
\neg \operatorname{Decor}(rid_i) \;\wedge\;
\operatorname{Sem}(c_i)
\Big].
\label{eq:filter}
\end{equation}
Here, $\operatorname{Sem}(c_i)$ indicates whether $c_i$ is semantically relevant, i.e., clickable, text-bearing, or of a common interactive type. Specifically, $click_i$ captures runtime clickability, whereas $\operatorname{IsInteractive}(\cdot)$ encodes prior knowledge of interactive widget classes when runtime flags are unreliable. In addition, $\tau_a$ is the minimum area threshold; $\operatorname{Decor}(\cdot)$ filters background and separator elements by decorative keywords or resource patterns; $\operatorname{HasText}(\cdot)$ checks whether readable text is present; and $\operatorname{IsInteractive}(\cdot)$ denotes common interactive widget types such as Button, EditText, and Switch. We further deduplicate text widgets representing different states of the same function using normalized keys.

\subsubsection{Adjacency Relations and Ordering}
To preserve consistency between the input sequence and spatial layout, we construct adjacency relations among retained widgets and derive an approximate reading order. For any two widgets $c_i$ and $c_j$, the distance between their centers is
\begin{equation}
d_{ij}=\left\|center(b_i)-center(b_j)\right\|_2.
\label{eq:dist}
\end{equation}
If $c_i$ and $c_j$ overlap in their horizontal or vertical projections and the gap between their edges is below a threshold, we connect them to form an adjacency graph $G=(V,E)$. Starting from the widget with the smallest $(y_1,x_1)$, we perform depth-first search on $G$ and prioritize nearer neighbors to obtain an ordered sequence
\begin{equation}
\pi=[c_{\pi(1)},\ldots,c_{\pi(m)}],
\label{eq:ordering}
\end{equation}
which is used in Stages III and IV.

Based on $\pi$, we serialize retained widgets into structured UI metadata. Let
\begin{equation}
\tilde{u}_i=(id_i,\;type_i,\;text_i,\;b_i,\;click_i,\;rid_i),
\label{eq:ui_item}
\end{equation}
denote the serialized representation of widget $c_i$. The structured UI metadata is then
\begin{equation}
\widetilde{U}=[\,\tilde{u}_{\pi(k)}\,]_{k=1}^{m}.
\label{eq:structured_meta}
\end{equation}
This metadata preserves the key attributes needed for subsequent marking and multimodal reasoning.\appnote{The complete component extraction and filtering procedure, including adjacency thresholds and DFS-based ordering, is detailed in Appendix~\ref{app:filtering}.}

\subsection{Stage III: SoM Construction and Precise Grounding}
In multi-window scenarios, defect identification requires both visual evidence and structural semantics. When regions are described only in natural language, multimodal models are prone to ambiguous references and localization errors. Set-of-Mark (SoM)~\cite{yang2023set} mitigates this problem by overlaying symbolic markers on key widgets and linking them to widget attributes, enabling precise reference by marker ID.
After obtaining the ordered key widgets, we construct a Set-of-Mark representation for precise grounding. Each retained widget is assigned a unique marker $m_i$, which is overlaid on the raw screenshot $I$ to generate a marked image $I^{\mathrm{SoM}}$. We also generate a mapping table from markers to widget attributes:
\begin{equation}
\mathcal{M}=\{(m_i,\; type_i,\; text_i,\; b_i,\; click_i)\}_{i=1}^{m}.
\label{eq:mapping}
\end{equation}
Together, the raw screenshot, the marked image, and the mapping table form the mark set. SoM converts ambiguous visual references into referable symbolic IDs, enabling the model to localize defect-related regions through $\{m_i\}$ and improving explanation verifiability.\appnote{An illustrative example of SoM construction is provided in Appendix~\ref{app:som}.}

\subsection{Stage IV: CoT-Constrained Multimodal Reasoning and Structured Reporting}
We use the marked screenshot $I^{\mathrm{SoM}}$ as visual input, and the serialized marker mapping $\mathcal{M}$ together with runtime context $\Gamma$ as textual input to a multimodal large language model. Formally, let
\begin{equation}
S=\operatorname{Serialize}(\mathcal{M},\Gamma),
\label{eq:text_input}
\end{equation}
then the model input is $\langle I^{\mathrm{SoM}},\;S\rangle$.

CoT prompting constrains the model to reason in three steps: 1) \textbf{structured interface understanding}: identify high-risk regions and key elements under the current window mode and layout; 2) \textbf{multimodal defect analysis}: jointly use the marked screenshot and structured UI metadata to verify candidate defect types, such as text overlap/truncation, widget occlusion, missing images, null display, conflicts with system UI, and split-screen/foldable incompatibility; and 3) \textbf{defect diagnosis and localization}: output the final defect type, associated SoM markers, and concise evidence and explanation.

The final output is a structured report:
\begin{equation}
\mathcal{R}=\{type,\; location,\; evidence,\; explanation\},
\label{eq:report}
\end{equation}
where $type$ is the defect category, $location$ is the set of SoM marker indices, $evidence$ contains verifiable observations such as overlapping text fragments, occlusion relations, or missing regions, and $explanation$ is a concise cause analysis. This output supports both defect inspection and subsequent statistical analysis.\appnote{A worked example of the multimodal chain-of-thought prompting process is provided in Appendix~\ref{app:cot}.}

\section{Experimental Setup}
\label{sec:exp_setup}

\noindent\textbf{Apps and scenarios.}
We evaluate the proposed method on 50 real-world Android applications from diverse categories, including social networking, entertainment, utilities, and education. Each application is tested under two settings to quantify the impact of window-mode changes on defect exposure and detection performance: (\emph{i}) conventional full-screen and (\emph{ii}) split-screen/foldable (abbreviated as split/fold). We evaluate the proposed method on 50 real-world Android applications from diverse categories, including social networking, entertainment, utilities, and education.\appnote{Representative defective applications, together with their tested versions, OS versions, devices, and download sources, are listed in Appendix~\ref{app:apps}.}

\noindent\textbf{Testing pipeline and baselines.}
%For each application, we execute the pipeline \emph{automated exploration and screenshot collection} $\rightarrow$ \emph{data preprocessing} (widget filtering + SoM) $\rightarrow$ \emph{defect reasoning and report generation}. On average, each app yields about 1{,}800 screenshots, and the end-to-end process takes about 2 hours, for a total runtime of about 100 hours. We compare our method with two representative baselines: OwlEye~\cite{liu2020owl}, a deep learning-based display defect detector, and YOLO~\cite{yolo11_ultralytics}, an object-detection-based visual baseline. For fairness, all methods use the same screenshots and follow the same annotation criteria, outputting defective screenshots and defect types when supported.
For each application, we first perform automated exploration and screenshot collection, then conduct data preprocessing, including widget filtering and SoM construction, and finally perform defect reasoning and report generation. On average, each app yields about 1{,}800 screenshots, and the end-to-end process takes about 2 hours, for a total runtime of about 100 hours. We compare our method with OwlEye~\cite{liu2020owl} and YOLO~\cite{yolo11_ultralytics}. For fairness, all methods use the same screenshots and follow the same annotation criteria. Our method is built on Qwen2.5-VL-32B and adapted via LoRA-based fine-tuning, with rank 8, alpha 16, and dropout 0. We train the model with AdamW for 100 epochs using a learning rate of $5\times10^{-5}$, a per-device batch size of 4, an effective batch size of 8, 4-bit quantization, double quantization, and gradient accumulation of 8.

\noindent\textbf{Evaluation protocol and metrics.}
We evaluate from three perspectives: scenario contribution, application-level reliability, and fine-grained defect-type performance. Specifically, we compare the numbers of exposed defects under the conventional and split/fold settings, evaluate whether a method identifies at least one defective screenshot for each application, and report Precision, Recall, and F1 for major defect types. At the application level, reliability is measured by the false positive rate (FPR) and false negative rate (FNR). Let $T(a)\in\{0,1\}$ be the human-annotated ground-truth label for application $a$, and let $P(a)\in\{0,1\}$ be the method prediction. Then,
\begin{equation}\label{eq:fpr_fnr}
\begin{aligned}
\mathrm{FPR} &=
\frac{\left|\{a \mid T(a)=0 \land P(a)=1\}\right|}
     {\left|\{a \mid T(a)=0\}\right|}, \\
\mathrm{FNR} &=
\frac{\left|\{a \mid T(a)=1 \land P(a)=0\}\right|}
     {\left|\{a \mid T(a)=1\}\right|}.
\end{aligned}
\end{equation}

% ===================== RESULTS =====================
\section{Results}
\label{sec:results}
\subsection{RQ1: Defect exposure under split-screen/foldable scenarios}

\begin{figure}[t]
  \centering
  \includegraphics[width=\columnwidth]{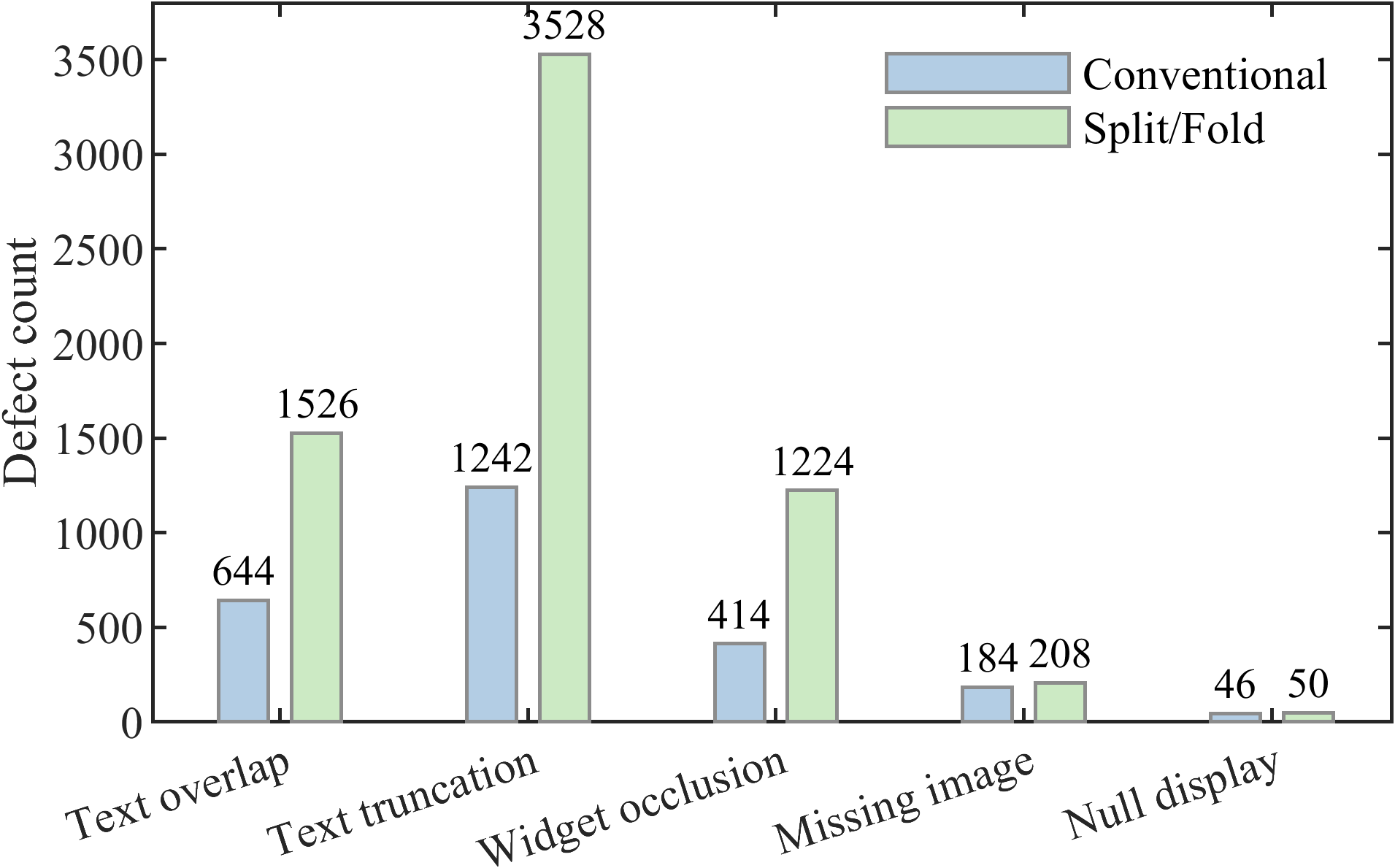}
  \caption{Triggered defect counts under conventional and split/fold settings.}
  \label{fig:rq1_defect_counts}
\end{figure}

We first examine whether split-screen and foldable operations expose more interface adaptation defects. For each application, we explore both the conventional and split/fold settings, collect screenshots, manually identify defective screenshots, and count defects by type. Fig.~\ref{fig:rq1_defect_counts} summarizes the results.

Split-screen/foldable settings substantially increase defect exposure, especially for layout-related defects. Compared with the conventional setting, text overlap increases from 644 to 1526 (+137\%), text truncation from 1242 to 3528 (+184\%), and widget occlusion from 414 to 1224 (+196\%). By contrast, missing image and null display increase only slightly. These results indicate that stronger spatial constraints and layout reflow amplify defects that are difficult to trigger in full-screen mode.

\subsection{RQ2: Coarse-grained comparison at the screenshot and app levels}

We next compare coarse-grained performance at the application level, without distinguishing defect types.

Table~\ref{tab:rq2_counts} reports the coarse-grained detection counts. Our method consistently detects more defective screenshots and identifies more defect apps than both baselines. Under the conventional setting, it detects 1503 defective screenshots and identifies 32 defect apps, compared with 1261/22 for OwlEye and 1334/24 for YOLO. Under split/fold settings, the advantage becomes larger: our method detects 5069 defective screenshots and identifies 40 defect apps, compared with 3567/27 and 4186/30 for OwlEye and YOLO, respectively.

\begin{table}[t]
\centering
% \footnotesize
\small
\setlength{\tabcolsep}{4pt}
\renewcommand{\arraystretch}{1.12}
\caption{Coarse-grained detection counts.}
\label{tab:rq2_counts}
\begin{tabular}{lcccc}
\toprule
\textbf{Method} &
\multicolumn{2}{c}{\textbf{Conventional}} &
\multicolumn{2}{c}{\textbf{Split/Fold}} \\
\cmidrule(lr){2-3}\cmidrule(lr){4-5}
& \makecell{\#screenshots}
& \makecell{\#apps}
& \makecell{\#screenshots}
& \makecell{\#apps} \\
\midrule
OwlEye        & 1261 & 22 & 3567 & 27 \\
YOLO          & 1334 & 24 & 4186 & 30 \\
\textbf{Ours} & \textbf{1503} & \textbf{32} & \textbf{5069} & \textbf{40} \\
\bottomrule
\end{tabular}
\end{table}

\begin{figure}[t]
  \centering
  \includegraphics[width=\columnwidth]{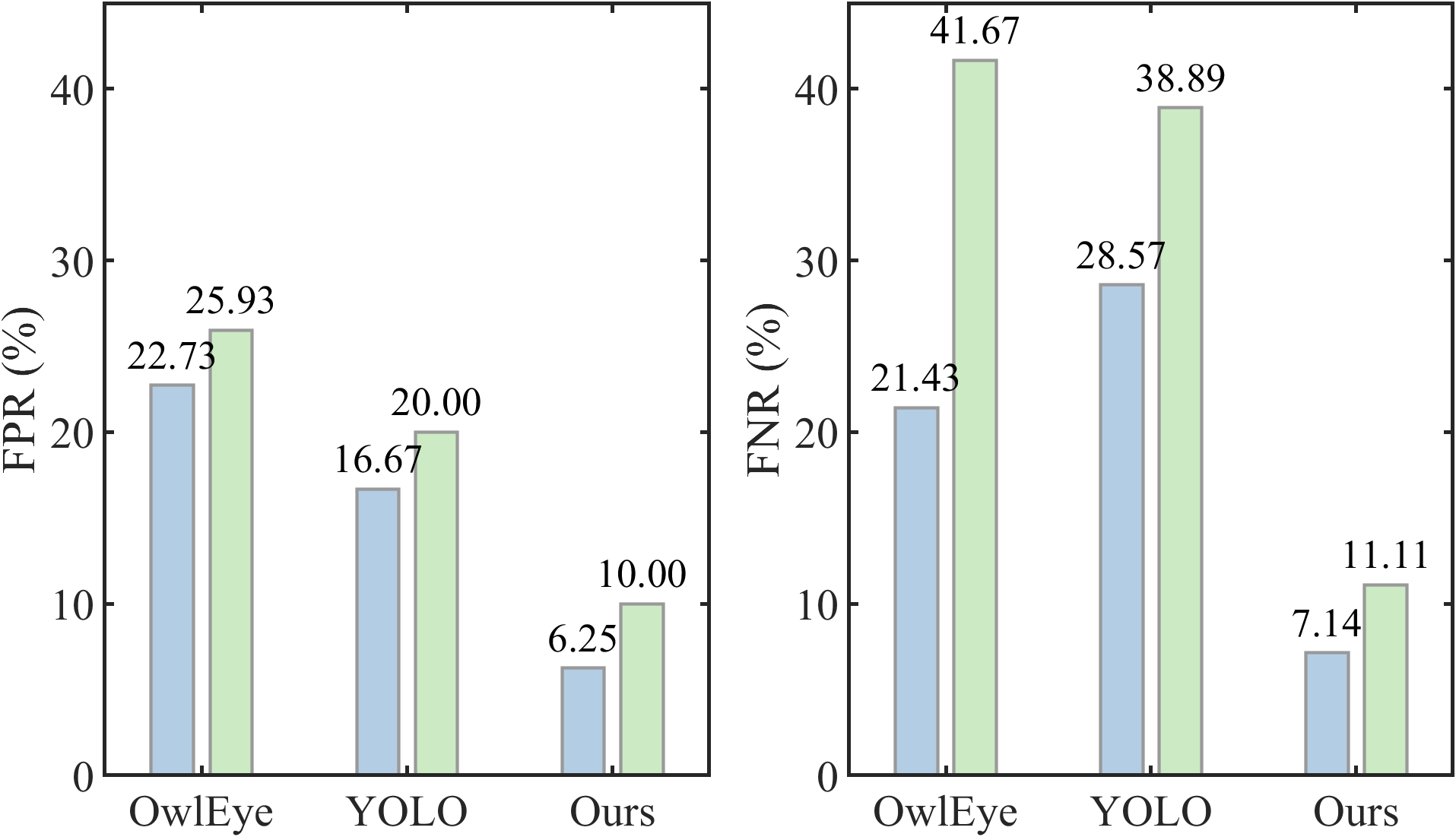}
  \caption{Application-level FPR and FNR under conventional and split/fold settings. Lower is better.}
  \label{fig:rq2_reliability}
\end{figure}

Fig.~\ref{fig:rq2_reliability} further compares application-level reliability using FPR and FNR. Our method achieves the lowest error rates in both settings. Under the conventional setting, its FPR/FNR are 6.25\%/7.14\%; under split/fold settings, they are 10.00\%/11.11\%, still clearly better than the baselines. This shows that our method not only detects more defective apps, but also provides better reliability.

\subsection{RQ3: Fine-grained analysis by defect types}

The third experiment evaluates defect-type-level performance. Table~\ref{tab:rq3_compact} compares detected instances across five defect types and quantifies the contribution of split/fold settings. Our method detects the largest number of instances for the five major traditional defect types, including text overlap, text truncation, widget occlusion, missing image, and null display. 
% It also identifies three additional defect types not covered by the baselines, namely app crash, foldable mismatch, and split mismatch. 
Moreover, text overlap, text truncation, and widget occlusion are mainly exposed under split/fold settings, each with a split share around 70\%.

\begin{table}[t]
\centering
\small
\setlength{\tabcolsep}{3pt}
\renewcommand{\arraystretch}{1.12}
\caption{Number of detected instances for each defect type, together with the corresponding split/fold counts and shares.}
\label{tab:rq3_compact}
\begin{tabular*}{\columnwidth}{@{\extracolsep{\fill}}lcccc@{}}
\toprule
\textbf{Defect type} & \textbf{Ours} & \textbf{OwlEye} & \textbf{YOLO} & \makecell[c]{\textbf{Split/Fold}\\\textbf{count (Ours)}} \\
\midrule
Text overlap     & \textbf{1228} & 887  & 995  & \textbf{859 (69.96\%)}  \\
Text truncation  & \textbf{2540} & 2186 & 2398 & \textbf{1803 (70.98\%)} \\
Widget occlusion & \textbf{1063} & 612  & 710  & \textbf{765 (71.97\%)}  \\
Missing image    & \textbf{192}  & 161  & 173  & \textbf{31 (16.15\%)}   \\
Null display     & \textbf{46}   & 39   & 41   & \textbf{8 (17.39\%)}    \\
\bottomrule
\end{tabular*}
\end{table}

Table~\ref{tab:rq3_prf} reports Precision, Recall, and F1 for the major defect types under split/fold settings. For \textbf{widget occlusion}, our method achieves the best Precision (87.6\%), Recall (86.8\%), and F1 (87.2\%), outperforming OwlEye and YOLO by 28.3 and 21.5 points in F1, respectively. For \textbf{text overlap}, it achieves the highest Recall (80.5\%) and best F1 (79.8\%), while YOLO attains slightly higher Precision (81.6\%). For \textbf{text truncation}, our method achieves the highest Recall (72.0\%), whereas OwlEye yields the highest Precision (82.1\%) and YOLO the best F1 (73.4\%), indicating a precision--recall trade-off. For the visually intuitive defects \textbf{missing image} and \textbf{null display}, all methods perform strongly, while our method achieves the highest Recall and F1 on both types.

\begin{table}[t]
\centering
% \scriptsize
\small
\setlength{\tabcolsep}{4pt}
\renewcommand{\arraystretch}{1.12}
\caption{Precision, Recall, and F1 scores (\%) under split/fold settings.}
\label{tab:rq3_prf}
\begin{tabular}{llccc}
\toprule
\textbf{Defect type} & \textbf{Method} & \textbf{Precision} & \textbf{Recall} & \textbf{F1} \\
\midrule
\multirow{3}{*}{Text overlap}
& OwlEye & 78.3 & 58.1 & 66.7 \\
& YOLO    & \textbf{81.6} & 65.2 & 72.5 \\
& Ours    & 79.2 & \textbf{80.5} & \textbf{79.8} \\
\midrule
\multirow{3}{*}{Text truncation}
& OwlEye & \textbf{82.1} & 62.0 & 70.6 \\
& YOLO    & 79.5 & 68.0 & \textbf{73.4} \\
& Ours    & 74.2 & \textbf{72.0} & 73.1 \\
\midrule
\multirow{3}{*}{Widget occlusion}
& OwlEye & 71.5 & 50.0 & 58.9 \\
& YOLO    & 75.8 & 58.0 & 65.7 \\
& Ours    & \textbf{87.6} & \textbf{86.8} & \textbf{87.2} \\
\midrule
\multirow{3}{*}{Missing image}
& OwlEye & 88.9 & 77.4 & 82.8 \\
& YOLO    & \textbf{92.1} & 83.2 & 87.4 \\
& Ours    & 89.7 & \textbf{92.3} & \textbf{90.9} \\
\midrule
\multirow{3}{*}{Null display}
& OwlEye & 89.7 & 78.0 & 83.4 \\
& YOLO    & \textbf{91.1} & 82.0 & 86.3 \\
& Ours    & 90.2 & \textbf{92.0} & \textbf{91.1} \\
\bottomrule
\end{tabular}
\end{table}

\section{Conclusion and Future Work}
This paper presents an end-to-end framework for interface display defect detection in split-screen, foldable, and other multi-window scenarios. By combining enhanced DroidBot-based scenario triggering, SoM-based visual grounding, and chain-of-thought-guided multimodal reasoning, the framework supports defect detection, localization, and explanation. Experiments on 50 real-world Android applications show that split/fold settings substantially increase the exposure of layout-related defects and that our method outperforms representative baselines in both application-level reliability and fine-grained defect detection, especially for semantically challenging defects such as widget occlusion.

Future work will extend the framework to more device types and adaptive layout scenarios, improve inference efficiency for large-scale testing, and incorporate richer interaction context to better detect defects arising during continuous user interaction.

\section*{Acknowledgment}
This work was partially supported by the National Natural Science Foundation of China (Grant No. 62502356) and the contract research grant from Dongfeng Motor Corporation (Project No. 202501hx1053; Project title: Development Project for a Personalized Driving System Based on End-to-End Autonomous Driving).

% ===================== REFERENCES =====================
%\balance
\bibliographystyle{IEEEtran}
\bibliography{ref}

\ifwithappendix
\clearpage
\appendix
\section{Additional Experimental Details}
This appendix provides additional implementation details, worked prompting examples, and representative defect cases used in our study.

\subsection{Examples of Collected Defective Applications}
\label{app:apps}

Table~\ref{tab:appendix_apps} lists example defective applications collected in our study, together with their tested versions, OS versions, devices, and download sources.

\begin{table}[!htt]
\centering
\footnotesize
\renewcommand{\arraystretch}{1.05}
\caption{Examples of collected defective applications.}
\label{tab:appendix_apps}
\resizebox{\columnwidth}{!}{%
\begin{tabular}{lllll}
\toprule
\textbf{Application} & \textbf{Tested Version} & \textbf{OS} & \textbf{Test Device} & \textbf{Source} \\
\midrule
MTG Familiar       & 3.6.6.dbg.3 & 8  & Google Nexus 6 & Google Play \\
Minesweeper        & 1.1 (2)     & 8  & Google Nexus 6 & Google Play \\
Signal-Android     & 7.40.2      & 12 & Xiaomi F22 Pro & Google Play \\
Super Productivity & 12.0.1      & 11 & Google Nexus 6 & F-Droid \\
Raccoon            & 1.14.0      & 12 & Samsung S10    & F-Droid \\
NewPipe-legacy     & 0.18.6      & 8  & Google Nexus 6 & GitHub \\
Dagger             & 1.2.0       & 8  & Google Nexus 6 & Google Play \\
Cards Score Keeper & 1.0.3       & 8  & Google Nexus 6 & Google Play \\
werewolf           & 1.0.1       & 8  & Google Nexus 6 & Google Play \\
Game Clock         & 1.0         & 8  & Google Nexus 6 & Google Play \\
sockstun           & 2.7         & 10 & Pixel 2        & GitHub \\
Orbot              & 17.4.1      & 15 & Google Pixel 8 & GitHub \\
Sky Map            & 1.9.6       & 8  & Google Nexus 6 & Google Play \\
toDoListb4a        & v2.0.0.3    & 11 & vivo v15       & GitHub \\
PocketMaps         & 3.7         & 8  & Google Nexus 6 & Google Play \\
metamask-mobile    & 7.43.0      & 10 & iPhone 15      & GitHub \\
My Position        & 1.3.5       & 8  & Google Nexus 6 & Google Play \\
Super Productivity & 12.0.1      & 11 & Samsung S10    & F-Droid \\
Metrodroid         & 3.0.0       & 8  & Google Nexus 6 & Google Play \\
\bottomrule
\end{tabular}%
}
\end{table}

\subsection{Defect Taxonomy and Fine-Tuning Data Distribution}
\label{app:taxonomy}

To support consistent labeling, model fine-tuning, and evaluation, we define a hierarchical defect taxonomy for GUI adaptation issues in split-screen and foldable scenarios. Based on the collected defect cases, we group the target defects into two major categories: \emph{missing/wrong information defects} and \emph{overlap/occlusion defects}. Figure~\ref{fig:appendix_taxonomy_examples} provides representative examples of the major defect categories in our taxonomy.

The first category, \emph{missing/wrong information defects}, refers to cases in which expected interface content is missing, incorrectly rendered, or displayed incompletely. This category includes \emph{missing image}, \emph{null display}, and \emph{text truncation}. Among them, text truncation is further divided into \emph{top truncation}, \emph{bottom truncation}, and \emph{side truncation}, according to which part of the text content is clipped by the layout.

The second category, \emph{overlap/occlusion defects}, refers to cases in which interface elements conflict spatially and one element partially or fully covers another. This category includes \emph{widget occlusion}, \emph{text overlap}, and \emph{navigation-bar/button overlap}. To better characterize different spatial conflicts, widget occlusion is further divided into \emph{widget-over-text}, \emph{text-over-widget}, and \emph{widget-over-widget} cases.

This taxonomy serves as the basis for constructing the fine-tuning dataset and organizing the evaluation results. It also helps distinguish genuine display defects from visually similar but acceptable design patterns, which is particularly important in complex adaptation scenarios where normal overlays and defective occlusions may appear similar at the pixel level.

\begin{figure}[!ht]
  \centering
  \includegraphics[width=\columnwidth]{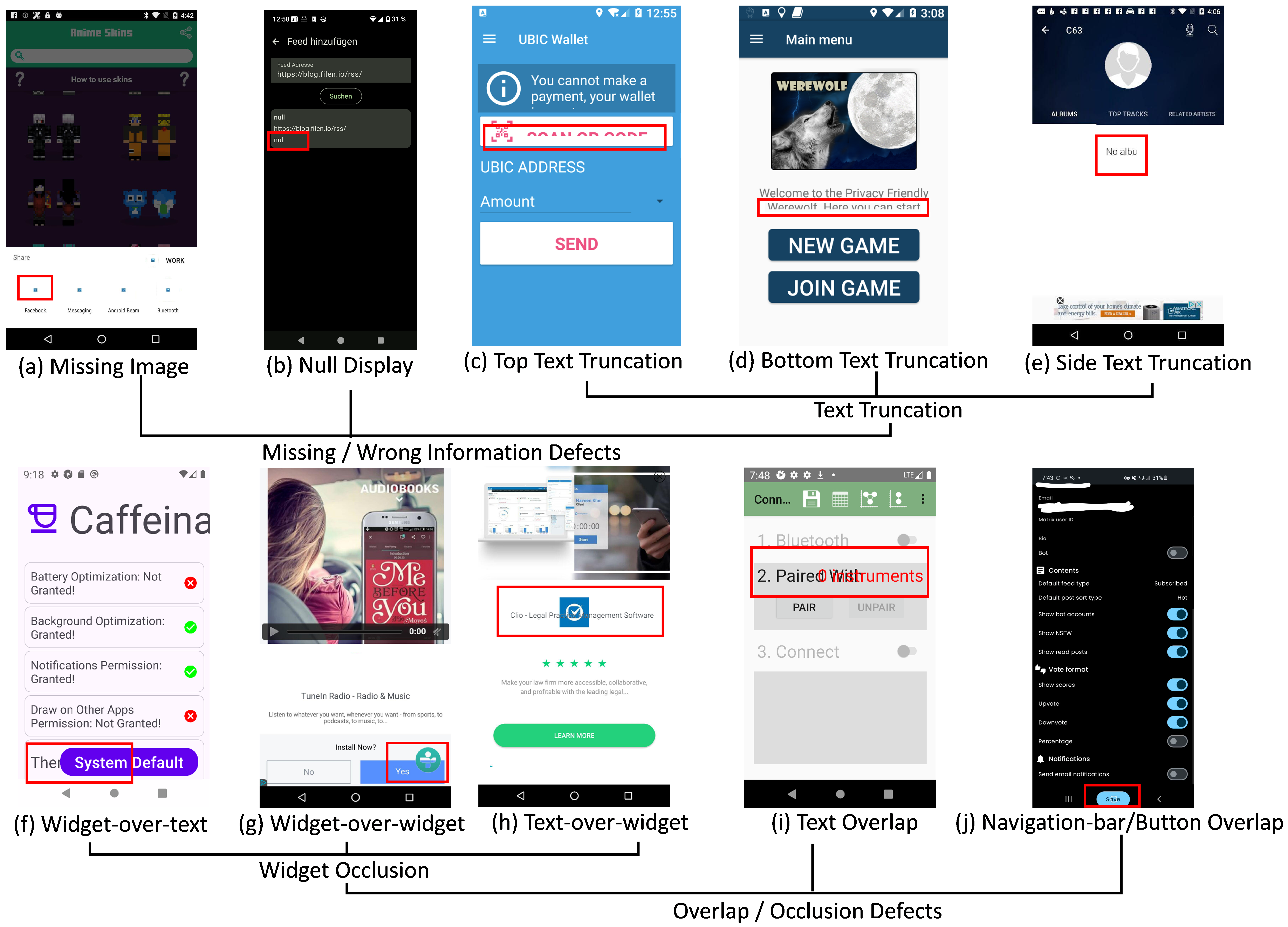}
  \caption{Representative examples of the major GUI adaptation defect categories in our taxonomy.}
  \label{fig:appendix_taxonomy_examples}
\end{figure}

Based on the taxonomy described above, we further refine the major defect categories into finer-grained subtypes to support defect recognition at different levels of granularity. In addition to the two primary GUI defect categories, the fine-tuning dataset also includes crash and compatibility-related cases, namely \emph{app crash}, \emph{split mismatch}, and \emph{foldable mismatch}. Moreover, the dataset contains normal-display samples, which help the model distinguish genuine GUI defects from visually similar but acceptable interface layouts. Figure~\ref{fig:appendix_window_examples} shows representative examples of normal and mismatched GUI layouts under foldable and split-screen scenarios.

\begin{figure}[!ht]
  \centering
  \includegraphics[width=\columnwidth]{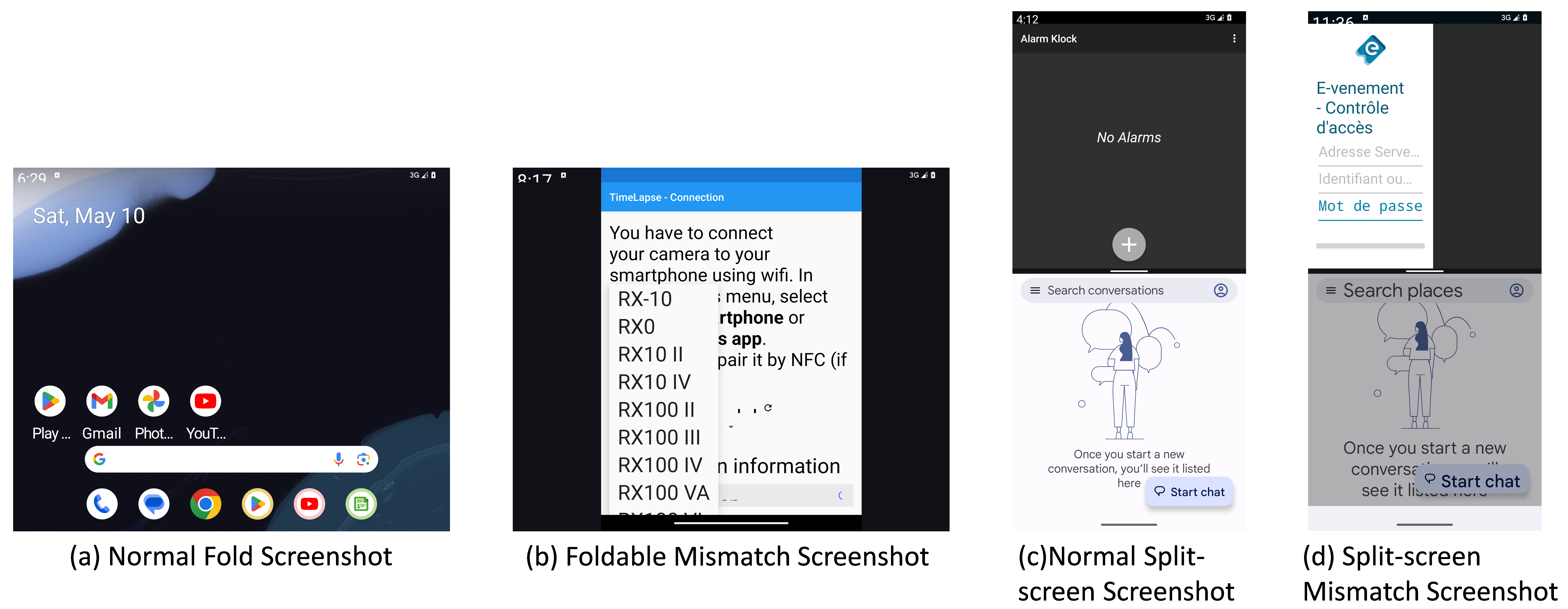}
  \caption{Examples of normal and mismatched GUI layouts under foldable and split-screen scenarios.}
  \label{fig:appendix_window_examples}
\end{figure}

Table~\ref{tab:appendix_ft_data} summarizes the distribution of the fine-tuning data used in our study.

\begin{table}[!t]
\centering
\footnotesize
\setlength{\tabcolsep}{4pt}
\renewcommand{\arraystretch}{1.1}
\caption{Distribution of the fine-tuning data.}
\label{tab:appendix_ft_data}
\resizebox{\columnwidth}{!}{%
\begin{tabular}{llll}
\toprule
\textbf{Major Category} & \textbf{Defect Group} & \textbf{Subtype} & \textbf{Count} \\
\midrule
\multirow{5}{*}{Missing / wrong information}
& --                & Missing image                 & 22 \\
& --                & Null display                  & 7  \\
& \multirow{3}{*}{Text truncation}
                    & Top truncation                & 7  \\
&                   & Bottom truncation             & 29 \\
&                   & Side truncation               & 6  \\
\midrule
\multirow{5}{*}{Overlap / occlusion}
& \multirow{3}{*}{Widget occlusion}
                    & Widget-over-text              & 10 \\
&                   & Widget-over-widget            & 6  \\
&                   & Text-over-widget              & 5  \\
& --                & Text overlap                  & 18 \\
& --                & Navigation-bar/button overlap & 2  \\
\midrule
\multirow{3}{*}{Crash / Compatibility}
& --                & App crash                     & 18 \\
& --                & Split mismatch                & 2  \\
& --                & Foldable mismatch             & 2  \\
\midrule
\multirow{1}{*}{Normal samples}
& --                & Normal display                & 30 \\
\bottomrule
\end{tabular}%
}
\end{table}

\subsection{Enhanced DroidBot Implementation Details}
\label{app:droidbot}

We extend DroidBot into an enhanced multi-window testing framework for collecting GUI adaptation data under split-screen and foldable scenarios. As shown in Fig.~\ref{fig:appendix_droidbot}, the framework adopts a modular architecture consisting of a multi-window controller, a GUI process monitoring module, and a parallel execution engine.

The framework supports four window-transition operations, namely \emph{Fold Test}, \emph{Split Test}, \emph{Drag Up}, and \emph{Drag Down}. These operations are used to trigger representative window-mode changes during automated exploration.

After each window operation, the framework records three types of raw data for subsequent analysis: application information, raw screenshots, and unprocessed widget information. Together, these outputs form standardized screenshot--UI-hierarchy--context triplets, which provide the basis for later SoM construction and multimodal defect reasoning.

\begin{figure}[!ht]
  \centering
  \includegraphics[width=\columnwidth]{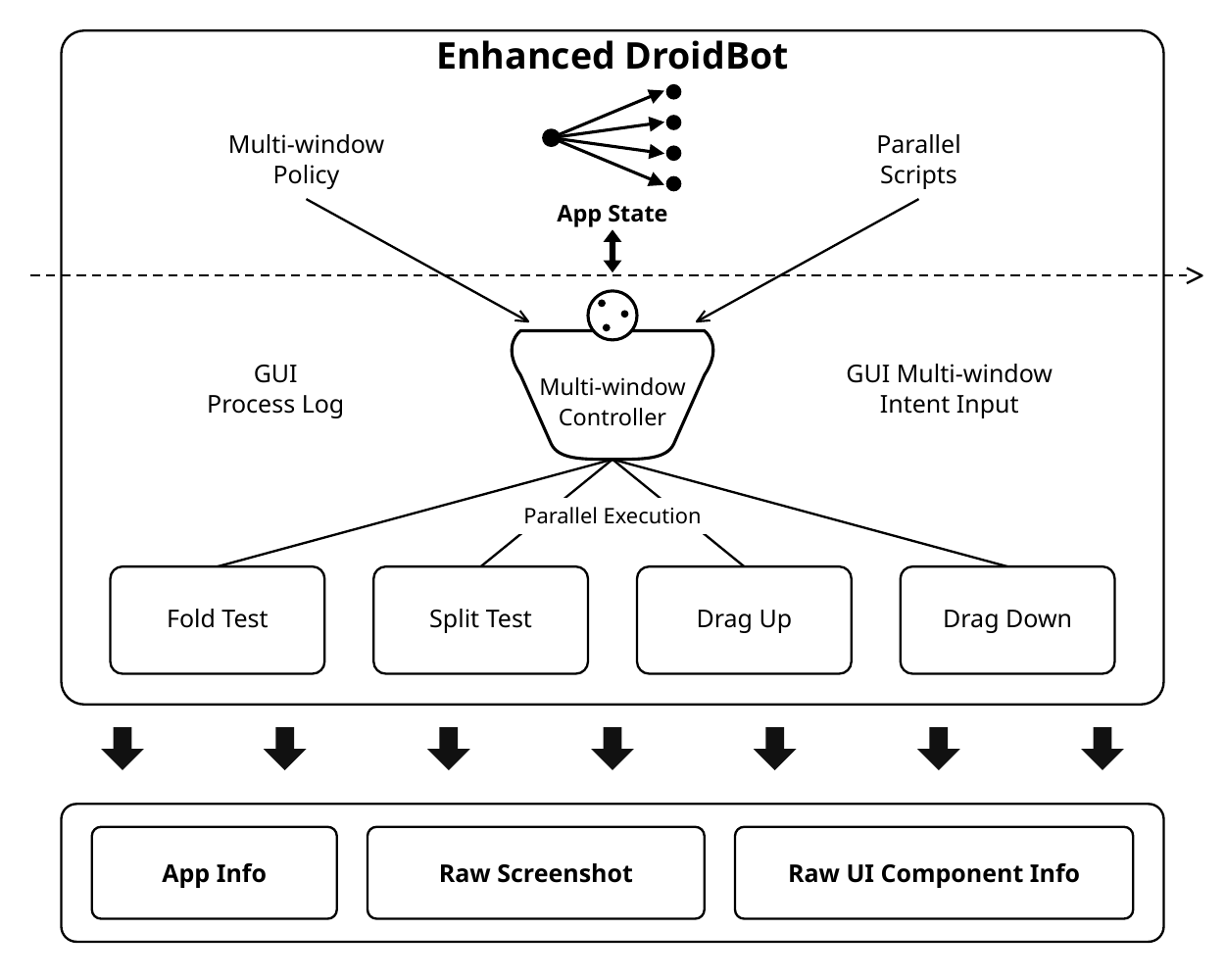}
  \caption{Workflow of the enhanced DroidBot framework for triggering split-screen and foldable interface states and collecting raw GUI data.}
  \label{fig:appendix_droidbot}
\end{figure}

\subsection{Component Extraction and Filtering}
\label{app:filtering}

To reduce noise in the raw UI hierarchy, we apply a component extraction and filtering procedure before constructing the final SoM representation. The procedure follows four core principles, namely size filtering, functional categorization, interactivity judgment, and content analysis. Its goal is to preserve components that are meaningful for GUI defect detection while minimizing the interference of redundant or irrelevant elements.

In our implementation, extremely small components are removed first. We then exclude decorative elements whose resource identifiers indicate background, divider, shadow, or similar visual-only functions. For container-type widgets, we retain them only when they are clickable or contain useful textual content. This design avoids passing large numbers of non-informative layout containers to the multimodal model.

After filtering, the retained components are connected through a spatial adjacency graph. Two components are considered adjacent only when they satisfy overlap constraints and their edge distance is below a direction-specific threshold. Following the original implementation, we use different threshold settings for the horizontal and vertical directions to better fit the layout characteristics of mobile interfaces. Neighboring components are then sorted by distance, and a depth-first traversal is used to generate an ordered component sequence that approximately follows the natural reading order of the interface. This process produces a more compact and semantically meaningful representation for downstream SoM construction and multimodal reasoning.

In our implementation, the horizontal adjacency threshold is configured with $(20,120,0.04)$, while the vertical threshold is configured with $(20,150,0.06)$, reflecting the different spatial characteristics of mobile GUI layouts. The DFS traversal starts from the top-left component after sorting controls by $(y_1, x_1)$, so that the resulting sequence better follows the natural reading order of the interface and preserves local functional groups.

Algorithm~\ref{alg:component_filtering} summarizes the complete extraction and filtering process.

\begin{algorithm}[!ht]
\caption{Component Extraction and Filtering}
\label{alg:component_filtering}
\small
\begin{algorithmic}[1]
\STATE \textbf{Input:} UI hierarchy JSON file $json\_file$
\STATE \textbf{Output:} Filtered component list $valid\_controls$, adjacency graph $adjacency\_graph$

\STATE Parse $json\_file$ and initialize $all\_controls \leftarrow \emptyset$
\FOR{each element $e$ in the UI hierarchy}
    \STATE Extract type, text, bounds, and attributes from $e$
    \STATE Construct a control object $c$
    \STATE Add $c$ to $all\_controls$
\ENDFOR

\STATE Initialize $valid\_controls \leftarrow \emptyset$
\FOR{each control $c$ in $all\_controls$}
    \IF{$c.width < 5$ \OR $c.height < 5$}
        \STATE \textbf{continue} \COMMENT{remove tiny elements}
    \ENDIF
    \IF{$c.resource\_id$ contains any of \{scrim, background, divider, shadow\}}
        \STATE \textbf{continue} \COMMENT{remove decorative elements}
    \ENDIF
    \IF{$c.type$ is a container widget}
        \IF{\textbf{not} ($c.clickable$ \OR $c$ has valid text)}
            \STATE \textbf{continue} \COMMENT{remove non-informative containers}
        \ENDIF
    \ENDIF
    \STATE Add $c$ to $valid\_controls$
\ENDFOR

\STATE Initialize $adjacency\_graph \leftarrow \emptyset$
\FOR{each pair of controls $(c_i, c_j)$ in $valid\_controls$}
    \STATE Check horizontal overlap and vertical overlap
    \STATE Compute horizontal edge distance $d_h$ and vertical edge distance $d_v$
    \IF{$(horizontal\_overlap \land d_h \le \tau_h)$ \OR $(vertical\_overlap \land d_v \le \tau_v)$}
        \STATE Compute Euclidean distance between the centers of $c_i$ and $c_j$
        \STATE Add an edge between $c_i$ and $c_j$ in $adjacency\_graph$
    \ENDIF
\ENDFOR

\FOR{each node $v$ in $adjacency\_graph$}
    \STATE Sort neighbors of $v$ by distance in ascending order
\ENDFOR

\STATE Sort $valid\_controls$ by $(y_1, x_1)$ from top to bottom and left to right
\STATE Select the top-left control as the DFS start node
\STATE Traverse $adjacency\_graph$ with DFS to obtain an ordered control sequence
\STATE Append isolated controls that remain unvisited
\STATE \textbf{return} $valid\_controls$, $adjacency\_graph$
\end{algorithmic}
\end{algorithm}

\subsection{SoM Construction}
\label{app:som}

After collecting screenshot--UI-hierarchy--context triplets with the enhanced DroidBot framework, we construct a Set-of-Mark (SoM) representation that aligns visual content with structured widget information. The goal of this step is to establish an explicit correspondence between interface regions and widget-level semantics, so that later multimodal reasoning can refer to specific interface elements more accurately.

Based on the retained widgets obtained through the component extraction and filtering procedure described in the previous subsection, we overlay widget bounding boxes and widget identifiers on the screenshot to form the final SoM representation. Different widget types are rendered with different marker styles, while widget identifiers provide explicit anchors for subsequent multimodal reasoning and localization.

\begin{figure}[!ht]
  \centering
  \begin{minipage}[t]{0.46\columnwidth}
    \centering
    \includegraphics[width=\linewidth]{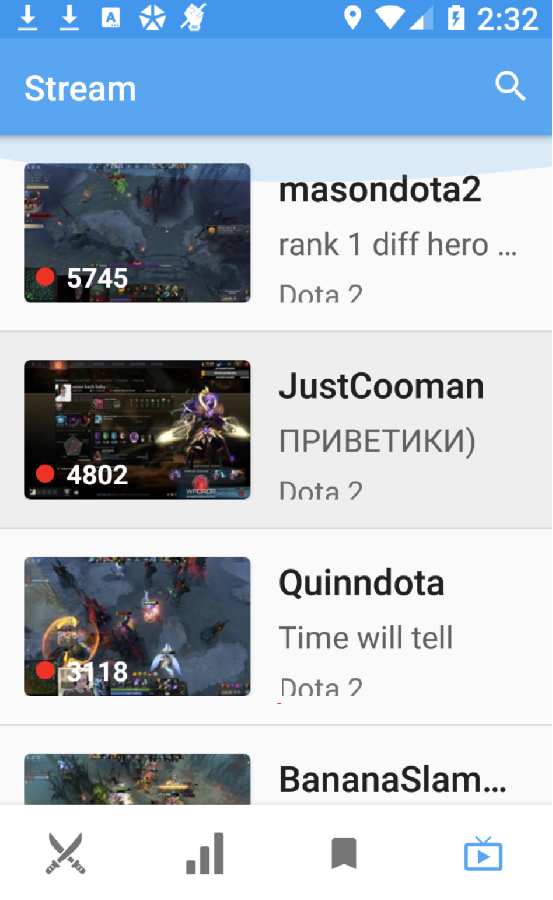}
    \vspace{2pt}
    \small (a) Original interface
  \end{minipage}
  \hfill
  \begin{minipage}[t]{0.46\columnwidth}
    \centering
    \includegraphics[width=\linewidth]{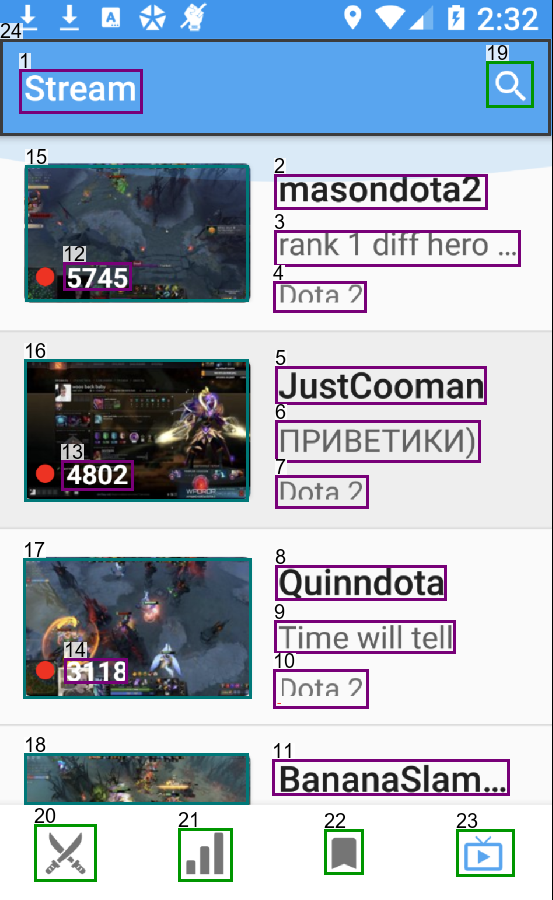}
    \vspace{2pt}
    \small (b) SoM-annotated interface
  \end{minipage}
  \caption{Example of SoM construction. The left image shows the original interface, and the right image shows the corresponding SoM-annotated interface.}
  \label{fig:appendix_som_example}
\end{figure}

\subsection{Multimodal Chain-of-Thought Prompting}
\label{app:cot}

To improve defect diagnosis in complex GUI adaptation scenarios, we design a multimodal chain-of-thought prompting framework that guides the model to jointly reason over the marked screenshot and the structured widget information. Instead of directly predicting a defect label from the image alone, the model is instructed to analyze the interface in a step-by-step manner, so that the intermediate observations remain consistent with the visual evidence and the final decision becomes more interpretable.

The prompting framework takes as input a screenshot annotated with SoM markers together with structured widget information, including widget types, coordinates, textual content, and widget identifiers. This design allows the model to simultaneously access visual appearance and interface structure, which is particularly important for diagnosing layout-related defects that depend on both semantic meaning and spatial relations.

The reasoning process is organized into multiple stages. The model first performs an overall inspection of the interface and identifies suspicious regions or abnormal layout patterns. It then conducts defect-specific analysis for the major defect categories considered in this work, including text overlap, widget occlusion, null display, text truncation, missing image, and navigation-bar/button overlap. In the final stage, the model combines visual observations with widget coordinates, widget types, and SoM identifiers to verify the suspected issue and produce the final defect decision.

This structured prompting strategy improves the consistency and interpretability of model predictions. Rather than producing only a defect label, the model is encouraged to output the defect type, the relevant widget or region, and a concise explanation of the diagnosis. Fig.~\ref{fig:appendix_cot} presents a worked example of the multimodal chain-of-thought prompting process, in which a text-overlap defect is identified through both visual and coordinate evidence.

\begin{figure}[!t]
  \centering
  \includegraphics[width=\columnwidth]{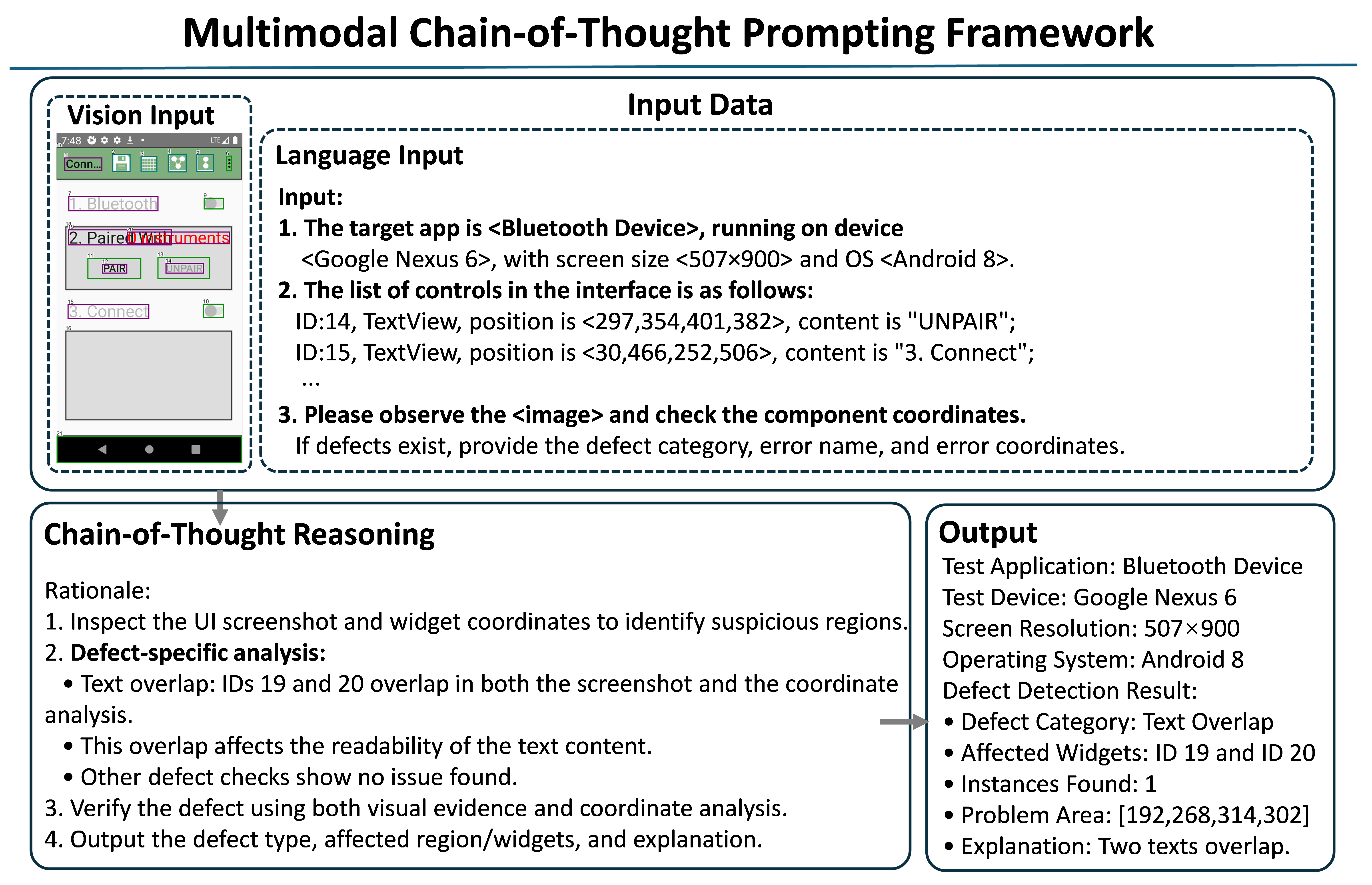}
  \caption{A worked example of the multimodal chain-of-thought prompting framework. The model takes a SoM-annotated screenshot and structured widget information as input, performs step-by-step defect reasoning, and outputs the defect type, affected widgets, problem area, and explanation.}
  \label{fig:appendix_cot}
\end{figure}
\fi

\end{document}